# EDUCATION AND ECONOMY – AN ANALYSIS OF STATISTICAL DATA

H.-U. Habermeier

Max-Planck-Institut für Festkörperforschung, Heisenbergstr. 1, D 70569 Stuttgart, Germany; *huh@fkf.mpg. de*

## ABSTRACT

In this paper the correlation between education, research and macroeconomic strength of countries at a global scale is analyzed on the basis of statistical data published by the UNIDO and OECD. It uses sets of composite indicators describing the economical performance and competitiveness as well as those relevant for human development, education, knowledge and technology achievement and correlates them. It turns out that for countries with a human development index (HDI) below 0.7 the basic education and technology achievement indices are the driving force for further development, whereas for the industrialized countries the knowledge index as a composite education and communication index has the strongest effect on the economic strength of a country as measured by the gross domestic product.

## INTRODUCTION

The role of education and its relation to economy is a topic of increasing public and political concern. Within the European Union e.g. a series of conferences with the title "The European House of Education: Education and Economy - a new Partnership" has been initiated in the late nineties. The responsible ministers or their deputies of 32 countries discuss this subject together with representatives of the business sector with the intention to pave the way for adjusting the education system to the needs of the society. The topics covered are concepts how to cope with the necessity of acquiring new qualifications, skills as well as further competence requirements for the information and knowledge based societies [1]. Rapid changes in the scientific, technological and economic environment in an increasingly interdependent world challenge existing educational systems worldwide. The consequences for education obtained via institutional systems such as schools and universities are seen in the necessity to get arranged with this steadily moving target. The inherent boundary conditions of these systems of being slow by nature make changes to meet the current requirements even more difficult. The concepts discussed controversially range from a new balance whether either personality development or skill acquisition dominated education should be at the core to a reinterpretation of the demands for internationality in terms of the necessity of language skills and mental openness to the different cultures of the economic players at the world level. Furthermore, education has to fulfill the task to teach people qualifications and skills which are in demand of the labor market. Access to timely, accurate, and diverse sources of knowledge is regarded as a major foundation of human development and as essential to stimulate economic growth, slow down population growth, build strong communities, and encourage democracy. Consequently, the acquisition of knowledge at all levels – primary, secondary as well as tertiary education schemes – is the essential ingredient to enable the population of a country to participate in the progress offered by the technological transformations of the 21st century. In this context, attempts to quantitatively measure the results of school education in the view of societal demands play an important role and enable the governments to focus on their comparative advantage/disadvantage at the level of countries and regions. Recently, several studies - like PISA [2] and TIMSS [3] - have been published, comparing the performance of students of different OECD countries after 4 years and 8 years of schooling, respectively. They result in an unexpected strength of some Asian countries (Korea, Japan), New Zealand, and North Europe (Finland). These studies have generated much political concern about the future of those countries with low or average marks. The consequence was a public debate about the importance of education as the base of any economically successful society. In that context, the role of education in natural science and math has been revisited and was regarded as the fertilizer for success in the increasingly interdependent world economy driven by innovation. The general discussion whether there is a causal connection between education and the level of macroeconomic performance of a country is much older than this debate. The correlation between economic strength, general education and education in science has frequently been postulated and these claims have been supported by a plethora of case studies[4,5,6]. The US National Center for Education Statistics (NCES) published 1997 an extensive study [7] based on an indicator analysis and linked the productivity of the US work force to the worker's education and skills. In this study, growth in education appears to be a substantial contributor to productivity growth, accounting for an estimated 10-20% of the growth of the gross domestic product [GDP] in the recent decades. Studies of comparative international economic growth consistently use education indices in their models to underpin the empirical findings. In labor economics e.g. it has been postulated that there is a substantial nexus between the performance of students, the quality of the educational system and economic competitiveness [8].

Complementary to these studies, focusing either on single elements of the complex relation of economy and education (e.g. the relation between internet use and economic performance) or the national aspects, only, in this paper an independent, more analytical - quantitative global approach is attempted. The aim of this study is a comparative investigation of indices describing the macroeconomic situation and perspectives of a country with those relevant for human development, education, and technological achievements and to find correlations between them. It is based on statistical data published by the OECD [9] and the United Nations Industrial Development Organization (UNIDO) [10]. Reliable correlations can only be found, if ensembles are studied large enough for statistics, therefore in this study the most recent data of all countries available in the reports are taken into account. In most cases the ensemble size was ~ 180 (i.e. the number of UN members), for some parameter sets the number was limited to 50 due to the lack of data at disposal. Based on these sets of data relations of the indices are derived and analyzed for industrial countries and developing countries, respectively. The paper is organized as follows: Subsequent to this introduction, Sect. 2 covers some comments on education as a general goal for personality development beyond direct knowledge acquisition and its direct impact on economy. The description of the method applied and some caveats using statistical data complement this part. In Sect. 3 the definitions of the indicators used are given, commonly taken as comparative measures to rank countries on a global scale. Subsequently, relations of economic indices with those relevant for education, science and technology are derived. The consequences drawn from these relations are given in Sect 4 for OECD countries and developing countries as well with special emphasis on the trends towards a knowledge society on a global scale.

## EDUCATION AS A HUMAN VALUE AND FUNDAMENTAL ELEMENT FOR PERSONALITY DEVELOPMENT

In a previous paper [11], the author developed some thoughts considering education as a general value beyond job qualification and skill acquisition and emphasized education in science – especially in materials science - as a fundamental element of personality development. That paper described a route for personality development based on materials science as pars pro toto, claiming that training in the different areas of materials science is much more than establishing merely a career path. The theses developed in that paper are:

1. Education, defined as the ability to live in the world with a balanced optimum of intellectual, emotional and sensual interaction, requires the awareness of and knowledge about the interaction of these three elements.
2. Materials determine the pace of development and the quality of life whose optimization requires an interdisciplinary scientific approach including economic and ecological issues. Consequently, natural science and humanities can not be treated independently and the barrier in between them has to be battered down.

3. Training in an interdisciplinary field such as materials science opens the mind of the persons involved, makes them aware of the interdependencies given in the total materials cycle and thus avoids the generation of narrow minded specialists.

In a subsequent study [12] the role of materials science for economic prosperity has been analyzed and consequences for materials science education have been derived. It stated that society can not benefit from the advances in science and technology without a continuous supply of well educated and well-trained scientists and engineers. To attract bright people and to get them excited for science is a task to be taken serious starting at the high school level. Education in science constantly faces new challenges arising from economical and ecological boundary conditions. The general problem of the exploitation of natural resources (esp. oil) for an increasingly developing world may serve as an example. In addition to the knowledge and skills that students should acquire during the conventional science classes, there are at least three more fields of teaching and learning necessary in order to be able to cope with the challenges due to globalization and the internationalization of science and technology:

- Knowledge of the basics of economics;
- Knowledge in history and the cultures of other countries, to meet the demands of internationality;
- Knowledge in improved communication skills, to bridge the gap between elites educated in natural science and in the humanities.

In this study the considered opinion is emphasized, that ability and knowledge acquisition beyond technical skills and job qualification are basic requirements to meet the challenges of the 21st century. To live and work constructively with full participation in all domains of life in an increasingly globalized world human abilities generally summarized as "emotional intelligence" are seen as mandatory.

The mutual understanding of boundary conditions set by different cultures and religions in various parts of the world is a prerequisite for conflict avoidance and conflict management. Many conflicts currently on the political agenda are arising from mutual miscomprehension of the cultural background of different parts in the world. Amongst them are the tension of the Western countries with the Islamic world as well as conflicts arising from the different concepts of human rights in the West (individualism based) and Far East (concept of the right of groups). These general remarks have to be kept in mind irrespective of the fact that this paper deals exclusively with the relation of the macroeconomic performance of countries with that based on education.

The method used in this contribution is the analysis of economy, education and technology related indicators as published in Refs. 9 and 10. In the field of macro-economy more than 400 indices for ensembles comprising 50-180 countries are available. It is obvious that a careful analysis of all of them in combination with the 200 education and technology related indices can not be accomplished in this paper. Furthermore, a dedicated selection of these combinations at random can reveal correlations that support any intention and/or prejudice. The famous Nobel-prize paradoxon [13] may serve as an example. It had been introduced by S. Kline to support the claim that there is a negative correlation between basic research and economic development. In Fig. 1 the data from Kline are plotted in a logarithmic scale for the average growth in the GDP showing that countries ranked high in Nobel prizes in science tend to be low in economic performance and vice versa. Keeping in mind that these statistics were done using small ensembles only, where statistical approaches are by no means justified, a simple re-plot of these data by changing the y-axis from the average growth rate to GDP or GDP per capita, reveals an opposite trend if any correlation at all (see Fig. 2). This example shows that neglecting elementary statistical rules and single element correlations can cause false conclusions.

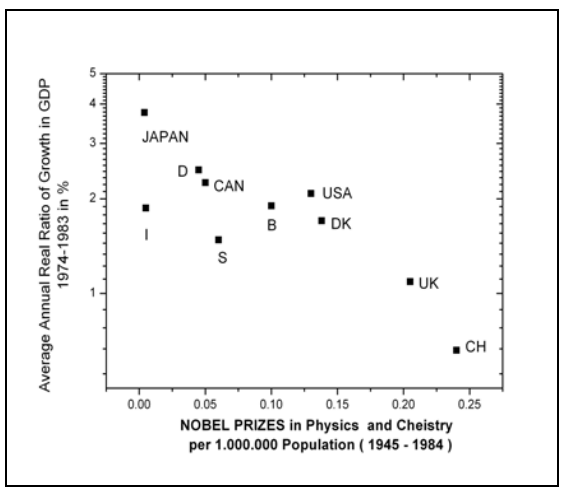

Fig. 1.  Growth in Gross Domestic Product and Nobel Prize-Winning in Physics and Chemistry.

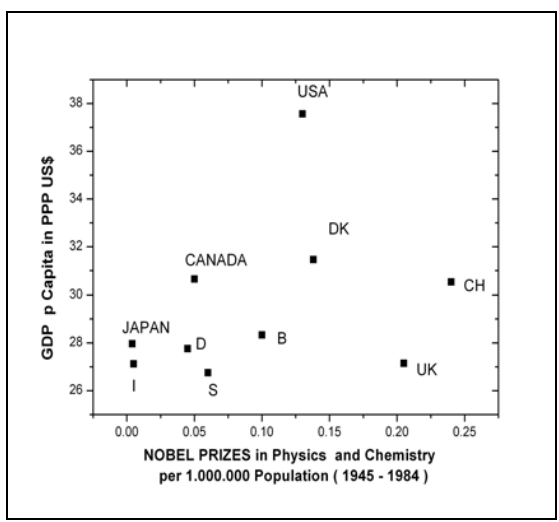

Fig. 2.  Gross Domestic Product and Nobel Prize-Winning in Physics and Chemistry .

In the search of reliable relations of macroeconomic and education/technology relevant indicators not only hard, measurable quantities such as GDP are used but also composite indicators combining various sets of relevant indices.  Here, the author takes those indicators only, that rest on a solid statistic base and are documented in various Human Development Reports and Industrial Development Reports of the UNIDO.

## DEFINITIONS OF THE RELEVANT INDICES AND THEIR RELATIONS

### Definition of the indices

*General Remarks*

A national economy provides people with basic necessities, luxuries, and opportunities for meaningful employment. The activities are usually grouped into production, services and administration. It represents a rather complex system that is internationally increasingly connected by the flow of goods, services, capital, and the working force with different degrees of restrictions. Due to this complexity, it is basically impossible to derive general conclusions from simple causal relations such as the link between the gross national product (GNP) per capita with one other parameter like expenditure on education or the adult literacy rate. Such sector analysis can reveal sectorial rather than general conclusions, only. To overcome these difficulties, the quantitative measures used in this study are mostly composite indices combining several criteria. Only those indices are used that are defined and listed in the Human Development Report of the UNDP, HDR [10], the Industrial Development Report 2005 and the relevant literature. According to the HDR, the definitions for index indicators, $I_V$, follow the common scheme

$$I_V = (V^{act} - V^{min})/(V^{max} - V^{min}) \qquad (1)$$

with $V^{act}$ being the actual observed value of a quantity (e.g. adult literacy or gross tertiary school enrolment) and $V^{max}$ and $V^{min}$ are the observed maximum and minimum values of the countries where data are available. Thus, the index indicators reach from 0 (lowest level) to 1 (highest level)

*Economic Indices*

To describe the strength of an economy, 3 index indicators are used in addition to the data for the GDP, reflecting the complexity of the economic environment. The directly measurable hard factors such as the gross domestic product (GDP) per capita in terms of purchasing power parity (GDPpC PPP) are used to measure the wealth of the population of a nation. Since the GDP as the only measure for the macroeconomic situation of a nation has several well known weaknesses ( contributions of the informal economy are missing, flaws due overexploiting natural resources etc.) some index indicators are used measuring in a complex way the economic competitiveness (ECI), the economic growth competitiveness (GCI), and the economic performance (EPI).

The economic competitiveness indicator (ECI) is used as defined in the World Competitiveness Yearbook (WCY) [14]. It analyzes and ranks the ability of nations to provide an environment that sustains the competitiveness of enterprises. The WCY divides the national environment into four main factors: (1) Economic performance (2) government efficiency (3) business efficiency and (4) infrastructure. Each of these factors is divided into 5 sub-factors highlighting every facet of the areas analyzed. These 20 sub-factors comprise more than 300 criteria. Criteria can be hard data, which analyze competitiveness as it can be measured (e.g. by the GDP), or soft data, which analyze competitiveness as it can be perceived (e.g. availability of competent managers). The detailed scheme to determine this index with the different weights of the sub-factors can be found in [14]. For 2003, 51 countries are covered. The highest ranks are 100 (for the US and Finland, the lowest is 9.8 for Venezuela. From these published data the ECI is determined according to eq. (1).

The growth competitive index (GCI) is composed of three component indexes: the technology index, the public institutions index, and the macroeconomic environment index. These indexes are calculated on the basis of both "hard data" and "survey data" according to:

Growth Competitiveness Index = 1/2 technology index + 1/4 public institutions index + 1/4 macroeconomic environment index. Here, the different sub indices enclose quite different areas ranging from hard facts such as number of patents, gross tertiary enrollment rate, relative R&D spending to problems of foreign direct investment in one country as an important source of new technology. The details of the determination of this index are given in Ref. 14. Again, the corresponding indicator is calculated according to eq. (1).

The economic performance index [EPI] is a sub-index of the ECI. It is derived from 90 criteria that provide a macro-economic evaluation of the domestic economy. These criteria include input data representing the domestic economy, international trade, international investment, employment and prices. 51 countries are covered including the 30 OECD members and 21 newly industrialized and emerging countries.

Even though these different indicators are not independent there is no simple relation between them. A plot of the GDPpC as of 2003 normalized to the value of the US ( $ 37.562 ) vs. ECI, EPI and GCI indicates that within a substantial scatter the GCI depends linearly on the GDP whereas the ECI and EPI have a trend for increasing GDP with increasing values of the composite indicators, only (see Fig.3).

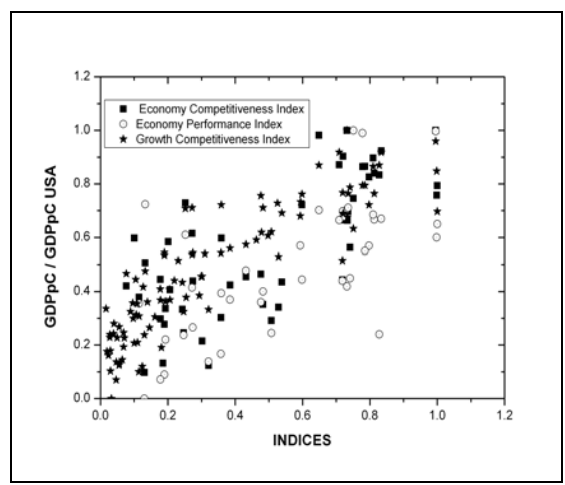

Fig. 3. Dependence of the gross domestic product in purchasing power parity normalized to the value for the USA on economy competitive indicator, economy performance indicator and growth competitiveness indicator.

### Human Development Index

The HDI measures the overall achievements in a country in three basic dimensions of human development - longevity, knowledge and a decent standard of living. It is measured by life expectancy, educational attainment (adult literacy and combined primary, secondary and tertiary enrollment) and adjusted income per capita in purchasing power parity (PPP) US dollars. The HDI is a summary, not a comprehensive measure of human development. As a result of refinements in the HDI methodology over time and changes in data series, the HDI can not be compared across different editions of the Human Development Reports. The data used in this paper are taken from the HDI Report 2005 and cover 177 countries with the highest value for Norway (HDI =0.963) and the lowest for Niger with a HDI of 0.281. Fig. 4 shows the dependence of the HDI index for these 177 countries on the GDP per capita. It indicates an exponential increase of the GDP per capita with the HDI with the highest value of 62.000 US $ for Luxembourg. Apparently, three regions in this plot can be identified, low HDI with a value below 0.5, a medium range with 0.5 < HDI > 0.8 and a range with HDI > 0.8. Amongst the countries with high HDI only a few have a GDPpC below 10.000 US$ (Costa Rica, Uruguay, Mexico, Bulgaria, Panama), the countries with medium HDI include China (rank 85) and India (rank 127), only a few such as Antigua (rank 60), Mauritius (rank 65), Oman (rank 71), Saudi Arabia (rank 77) and South Africa (rank 120) have a GDP per capita above 10.000 US$. Amongst the low HDI countries are the majority of the African countries with values for the GDPpC below 5.000 US$ (Swaziland, rank 147).

### Education Index (EI) and Knowledge Index (KI)

The education index measures a country's relative achievement in both adult literacy and combined primary, secondary and tertiary gross enrollment. First, an index for adult literacy and one for combined gross enrollment are calculated. Then the two indices are combined with 2/3 weight given to adult literacy and 1/3 to the combined gross enrollment. A higher value indicates a higher level of education. The EI emphasizes basic skills (adult literacy) and is certainly a measure more appropriate to differentiate the education level of developing rather than of industrialized countries where the EI is close to 1 for all OECD countries. The coverage here is 173 countries with an EI of 0.9-0.99 for the OECD states and the lowest of 0.33 for Sierra Leone.    .

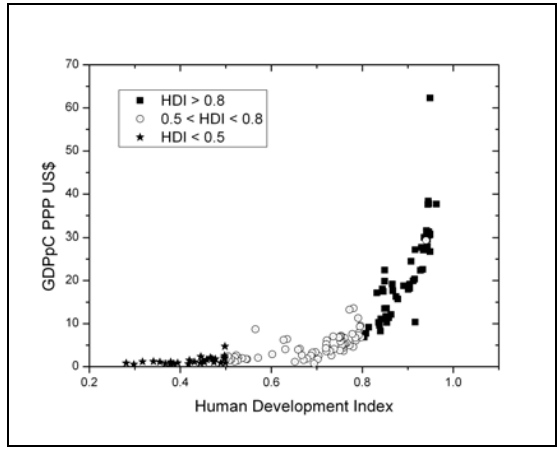

Fig. 4.  Relation of the GDP per capita in purchasing power parity (US $) and the human development index.

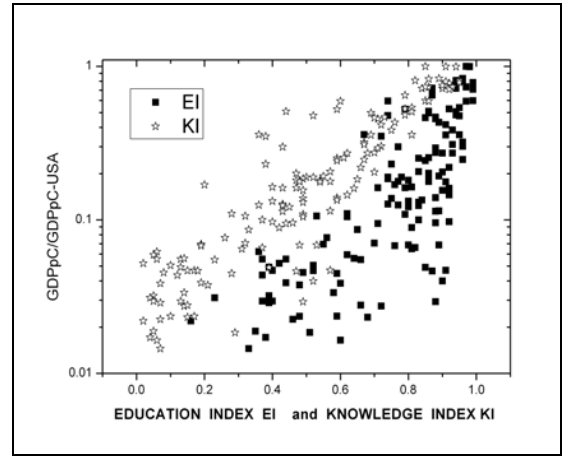

Fig. 5.  Relation of the normalized GDP per capita and the education index and knowledge index.

The knowledge index is the weighted average of the education index (primary and secondary school enrolment) and the communication index (telephone and internet access) with 2/3 weight on the education index and 1/3 on the communication index, respectively. The data used cover 180 countries and the years mid 1990 to late 1990. For the general discussion a further breakdown into more specific parameters such as percentage of students in primary, secondary and tertiary enrollment or expenditure pre student for the different levels will be done in a forthcoming paper and is partially done in Sect. 4. In Fig. 5 available data for the EI and KI for all countries are represented, showing the clear exponential trend especially for the knowledge index. The deviations from the trend are mainly caused by countries with a relatively high education index, however, being not so successful in their economy. They include African countries like Nigeria and the Congo as well as Moldavia, Kyrgyzstan and Tajikistan (lower right corner of Fig. 5).

*Technology Achievement Index (TAI) and Technology Index (TI)*

The TAI is a composite index of technological achievements of a country reflecting the level of technological progress and thus the capacity of a country to participate in the network age. The index aims to capture technological achievements of a country in four dimensions:

- Technology creation, as measured by the number of patents granted to residents per capita and by receipts of royalties and license fees from abroad per capita (TC),
- Diffusion of recent innovations, as measured by the number of internet hosts per capita and the share of high- and medium technology exports in total goods exports (DrI)
- Diffusion of old innovation as measured by telephones per capita and electricity consumption per capita (DoI),
- Human skills as measured by mean years of schooling in the population aged 15 and above and the gross tertiary science enrolment ratio (HS).

For each of the indicators in these dimensions the respective indicator indices are calculated according to equ. (1) And the TAI is determined as the simple average

TAI = ¼(TC +DrI + DoI + HS)

A higher value indicates greater technological achievement.

The TAI focuses more on outcome and achievements rather than on efforts such as numbers of scientists or R+D expenditures. It reflects how well a country as a whole is participating in creating and using technology. It is not a measure of heading global technology development.

The country coverage is 72 with data from 1995-2000, the highest rank is 0.74 for Finland, the lowest 0.07 for Sudan.

The technology index measures the level of technological advancement as the equal combination of three quantities:

- The number of computers with active Internet Protocol addresses connected to the internet per 10,000 people
- The number of telephone lines connecting a customer's equipment to the public switching telephone network per 10,000 people
- The number of people using portable telephones and subscribing to mobile telephone services per 10,000 people. The country coverage as of 2002 is 198 with the highest value of 1 for Finland and the lowest of 0 for the Democratic Republic of Congo.

Whereas the TAI measures a broader field ranging from innovations to enrollment in tertiary education, the TI is restricted just to achievements in communication. The relation of the two is given in Fig. 6 indicating a rather good fit with a quadratic relation.

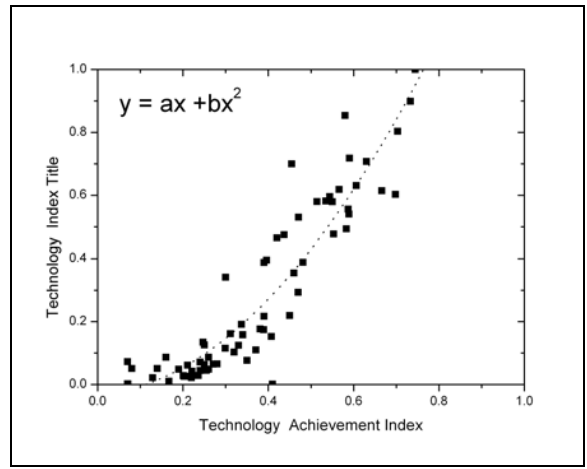

Fig. 6.  Relation of the technology index with the technology achievement index.

**The relation of economic indicators with education and technology indicators**

In Fig. 5 the data for the GDPpC normalized to the value for the US are presented versus the education index (EI) and the knowledge index (KI) in a semi logarithmic plot. The main trend of the GDP with the indicators can easily be seen irrespective of the large scatter of the data. Both indices can be approximated by an exponential dependence of the type

GDP/ (GDP-US) ~ Aexp (α I)            (2)

With I being the indices EI and KI, respectively. The prefactor in the exponent is for the education index α = 1.95 and somewhat smaller for the knowledge index (α = 1.64); i.e. the increase of the GDP is stronger for the EI compared to the KI.

The attempt to find correlations of the macroeconomic indicators EPI, ECI and GCI with education and knowledge related indices revealed a trend of increasing economic performance and competitiveness with increasing EI and KI. Due to the large scatter of the data no further analysis of the data was made. It is obvious that the education index with its special emphasis on adult literacy is not suitable for correlations with the rather complex economic indicators.

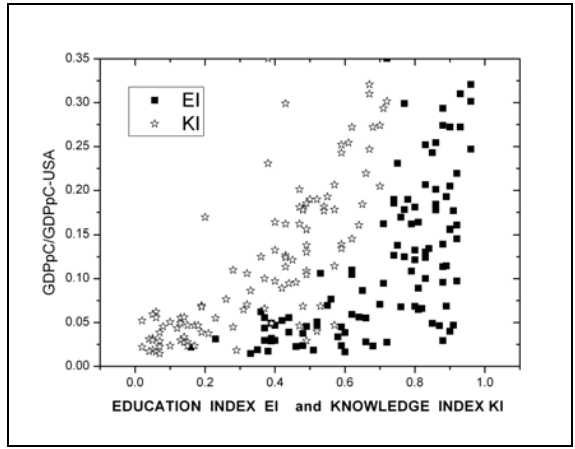

Fig.7. Relation of the GDP per capita normalized to the value for the US and the education index and knowledge index for countries with low GDP per capita.

Taking Fig. 3 it is clear that the analysis should be made separately for industrialized countries and developing countries. The division here is done rather arbitrarily by defining the border line at 30% of the GDPpC of the US value. This definition, however, matches roughly with the classification of countries with high human development (HDI > 0.8) as industrialized countries in the HDI reports.

For developing countries (normalized GDPpC < 30% of US) the dependence of the GDP on the education index and knowledge index, respectively shows a striking functional similarity as shown in Fig. 7. Since the education index includes data for the adult literacy by a weight of 2/3, the higher values for the education index can easily be explained. Extracting the contribution due to adult literacy from the education index, the revised data coincide with the values for the knowledge index within the scatter. The data can be described by an exponential behavior according to eq. (2) With a factor $\alpha$= 1.3. Similarly, the corresponding dependences of the GDP on the technology index and the technology achievement index yielded much higher values for $\alpha$ of 2.2 and 2.78, respectively. The consequence to be drawn from this is that driving force for the GDP increase has to be seen in the contributions entering the TAI and

TI. Attempts to correlate the macroeconomic indicators EPI, ECI and GCI with education related indices failed due to the large scatter and the small ensembles of data available.

For the industrialized countries (GDPpC > 30% of that of the US) the picture emerges to be quite different. First, due to the small variation of the education index within that ensemble (0.86<EI<0.99) the education index can not be accounted as the source of any macroeconomic differences between the countries. The indices able to describe the dependence of the GDPpC are mainly the knowledge index, the technology achievement index and the technology index. Fig. 8 shows the log GDPpC/GDPpC-US as a function of these three indices. Even with the scatter of the data the coefficients in the exponential relation can be derived as $\alpha = 0.56$, 0.9 and 1.4 for the TI, TAI and KI, respectively. This reveals clearly the important role of the knowledge index and emphasizes the importance of primary and secondary school enrollment in combination with communication skills such as internet use.

Furthermore, performing the similar analysis for the economic competitiveness indicator, ECI, and the growth competitiveness indicator, GCI, the value for $\alpha$ in all cases is largest for the KI and smallest for the TI. The set of data for α is given in Table 1 for the GCI and the ECI. The analysis of the economic performance indicator yielded no clear correlations. The values for α given in Table 1 clearly show, that the combination of school enrollment (primary and secondary) and communication skills determines the pace of the development more that the technology index measuring mainly communication issues (internet and phone use). Comparing the GDP related data for developing and industrial countries it can be clearly seen that the pace for economic improvement is higher in developing countries compared to industrialized countries.  Fig. 9 shows this for the three indices TI, TAI and KI.

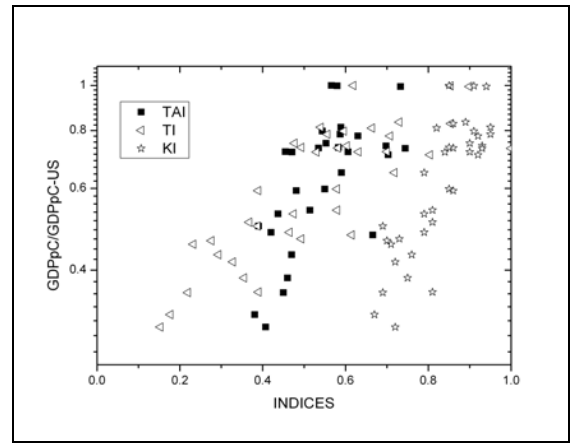

Fig.8. Relation of the GDP per capita normalized to the value for the US and the knowledge, technology and technology achievement indices for countries with a GDP per capita > 30% that of the US.

Table 1.    Prefactors of the exponent in the dependence of ECI and GCI on the KI, TAI and TI, respectively.

|  | Economic Competitiveness Indicator | Growth Competitiveness Indicator |
|---|---|---|
| Knowledge Index | 2.2 ± 0.5 | 0.92 ±0.2 |
| Technology Achievement Index | 1.5 ± 0.3 | 0.82  ± 0.2 |
| Technology Index | 0.86 ± 0.2 | 0.4 ± 0.1 |

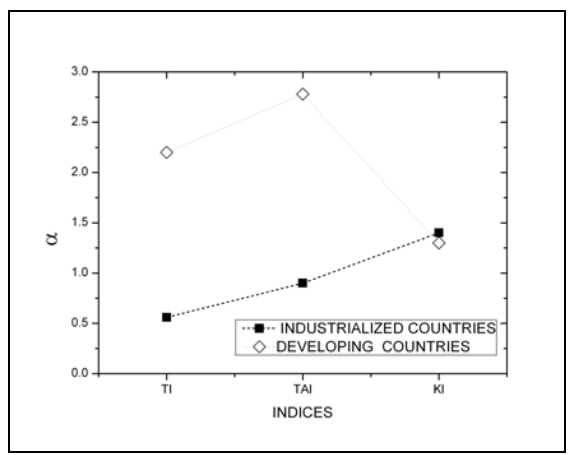

Fig. 9. Exponent in GDP growth on technology, technology achievement and knowledge indices for developing and industrialized countries.

In Fig.9 two basic messages are buried. First, the value of α that determines the speed of economic developments is larger for developing countries compared to industrialized nations. That is the consequence of the fact, that a growth in the GDP of 5% for a developing country such as Algeria (GDP PPP US$ for 2003 is 0.066 bn $) yields 0.0033 bn $ increase which is 1/3000 of that of Poland for the same growth rate and an equal population (~ 30 million).  Second, it can be concluded that the most important driving force for economic progress for the developing countries is the index for technology achievement whereas for the industrialized countries the knowledge index plays the leading role.

## CONCLUSIONS

In order to draw further conclusions from these empirical findings, a closer look at the components of the TAI and KI is required.  In both indices education – as measured in years of schooling or percentage of tertiary education enrollment – plays a substantial role as outlined in Sect. III. The quality of education, however has not been taken into account, yet. Two indicators can be used to overcome this shortcoming. One is the consideration of the outcome of the TIMSS and PISA studies reflecting on mathematical and scientific literacy and a combination of both with reading literacy, the other is dealing with expenditures for education as a whole, and the expenditure per student at the different levels. Additionally the expenditures on R&D in addition to education can shed some light on the economic development of countries. The data used are mainly from the OECD countries, i.e. restricted to ~ 30. Further analysis is done by searching for correlations between the GDP as economy indicator the knowledge index and technology achievement indices as the driving forces for development and the TIMSS and PISA data. Relating the GDP per capita with the normalized PISA and TIMSS performance we find again an exponential increase of the GDPpC with increasing performance. In Fig. 10 the data are given for the combined reading, mathematical and scientific literacy after 8 years of schooling. Both sets of data coincide, indicating that scientific literacy plays an important role for the economic development of a country.

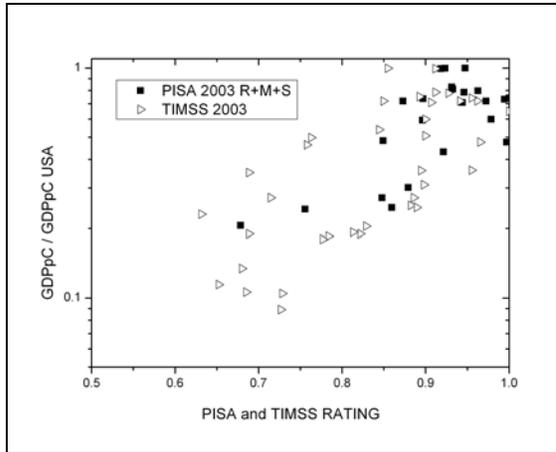

Fig. 10. Relation of the normalized GDP per capita and the TIMSS and PISA ratings

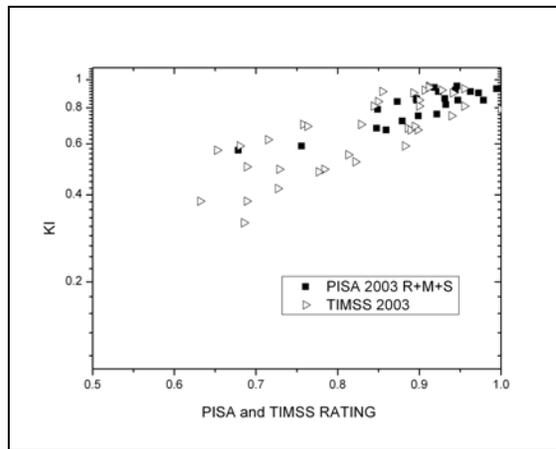

Fig.11. Dependence of the knowledge index on the PISA and TIMSS ratings

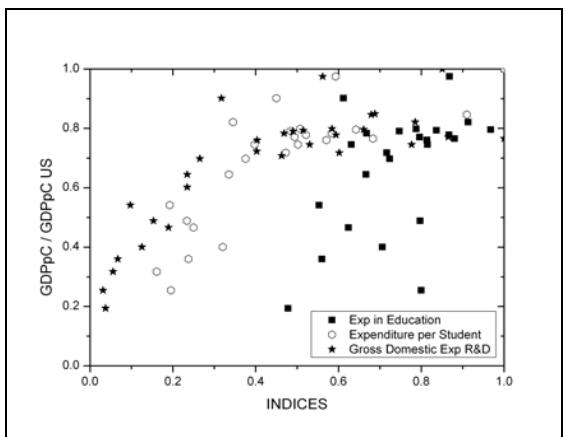

Fig. 12. Dependence of the normalized GDP on the expenditures on education, expenditure per student in the tertiary education and the gross domestic expenditures in % of GDP. The data are normalized to the maximum values (see text).

For industrialized countries it has been revealed that the knowledge index is the leading measure for economic growth. Its dependence on the education performance as revealed from the TIMSS and PISA studies is shown in Fig. 11, indicating an exponential increase of the knowledge index with the ratings. The obvious conclusion is that an improvement of education giving rise to a higher knowledge index will be substantial for economic improvement.

Finally, some financial aspects of education are addressed. Fig. 12 shows the relation of the GDP per capita normalized to the US and the normalized values for the tertiary education expenditure per student (normalized to the US value of 22.200 $), the expenditure in education in % of the GDP (normalized to US with 0.89% 0f GDP for public and private education) and the gross domestic expenditure in R&D (normalized to the highest value – Sweden with 0.97% of GDP). Whereas for the expenditures on education – as measured in % of GDP - no clear trend is observed, the expenditures per student in the tertiary education and the gross domestic expenditures on R&D (GERD) can be described by

$$GDPpC/GDPpC\text{-}USA = \beta \, (\, INDEX)^{1/2}$$

With nearly identical values $\beta = 0.83$. This implies that for the most industrialized countries with high GDP a simple doubling of expenses does not yield a doubling in the effect on the GDP.

In conclusion, the analysis shows, that for developing countries investment in basic education and technology achievement are the driving forces for the improvement of human development. For industrialized countries the main emphasis must be put on the improvement of the knowledge base. Here, the almost identical relations of the normalized GDP per capita and the education performance as revealed by TIMSS and PISA show that mathematical and science literacy are the most important contributions. The education in science therefore is a mandatory prerequisite for sustainable economic performance of a country.

## REFERENCES

1.  Conference of the European Ministers of Education,  June 24th, 1999, Budapest
2.  Learning for Tomorrow's World – First Results from PISA 2003, OECD,
3.  M.O. Martin, TIMSS 2003, *www.timss.bc.edu*
4.  J. Adams, Journal of Political Economy **98,** 673 (1990)
5.  J.J. Furman, M.E. Porter, and S. Stern, *Research Policy* **31**, 899 (2002)
6.  C. Jones, *J. of Political Economy* **103**, 739 (1995)
7.  P.T. Decker, J.K. Rice, M.T. Moore and M.R. Rollefson, US Department of Education, NCES 97-269, Washington DC (1997 )
8.  M. Bils and P. Klenow, *The American Economic Review* **90**, 1160 (2000)
9.  OECD in Figures 2004, ed. R. Clarke, *www.oecd.org/bookshop*
10. Human Development Report 2003, UNDP, Published by Oxford University Press, Inc. New York, N.Y., USA, 2003
11. H.-U. Habermeier, *Mat. Sci. Eng.* **A 199,** 99 (1995)
12. H.-U. Habermeier, *J. of Materials Education* **24**, 87 (2002)
13. *World Competitiveness Yearbook 2003*, ed. S. Garelli,  Institute for Management Development, Lausanne 2003.